# 3D structured Bessel beam polarization and its application to imprint chiral optical properties in silica

Jiafeng Lu,[1,2] Mostafa Hassan,[3] François Courvoisier,[3] Enrique Garcia-Caurel,[4] François Brisset,[1] Razvigor Ossikovski,[4] Xianglong Zeng,[2] Bertrand Poumellec,[1] and Matthieu Lancry[1,*]

AFFILIATIONS:

[1] Institut de Chimie Moléculaire et des Matériaux d'Orsay, Université Paris Saclay, Orsay 91405, France
[2] Key Laboratory of Specialty Fiber Optics and Optical Access Networks, Joint International Research Laboratory of Specialty Fiber Optics and Advanced Communication, Shanghai Institute for Advanced Communication and Data Science, Shanghai University, Shanghai 200444, China
[3] FEMTO-ST Institute, University of Franche-Comté, CNRS, Besançon 25030, France
[4] LPICM, CNRS, Ecole Polytechnique, Institut Polytechnique de Paris, Palaiseau 91128, France
[*] Corresponding author: matthieu.lancry@universite-paris-saclay.fr

**Polarization plays crucial role in light-matter interactions; hence its overall manipulation is an essential key to unlock the versatility of light manufacturing, especially in femtosecond laser direct writing. Existing polarization-shaping techniques, however, only focus on the manipulation in transverse plane of a light beam, namely a two-dimensional control. In this paper, we propose a novel passive strategy that exploits a class of femtosecond laser written space varying birefringent elements, to shape the polarization state along the optical path. As a demonstration, we generate a three-dimensional structured Bessel beam whose linear polarization state is slowly evolving along the focus (typ. 90° within 60λ). Such a "helical polarized" Bessel beam allows imprinting "twisted nanogratings" in $SiO_2$ resulting in an extrinsic optical chirality at a micrometric scale, which owns a high optical rotation. Our work brings new perspectives for three-dimensional polarization manipulations and would find applications in structured light, light-matter interaction and chiral device fabrication.**

## I. INTRODUCTION

Polarization, as a fundamental property of light, lies in the core of many important light applications in various domains, such as light-matter interaction [1], optical display [2,3], light sensing [4], and data storage [5]. In light science, polarization refers to the vibration mode of the electric field vector of light and thus, corresponds to the spin angular momentum of the photon. Hence, polarization manipulation can control the light-matter interaction in a desirable manner, to unlock the versatile functions of light-driven technologies, especially in ultrafast laser manufacturing.

Light-matter interaction at the femtosecond (fs) scale consists in the absorption through nonlinear photoionization mechanism, which enables precise energy deposition in any volume of the transparent materials without surrounding damage [6]. Such a mechanism results in diverse modifications inside the material that are dependent on the laser parameters, such as refractive index three-dimensional (3D) profiling [7], anisotropic subwavelength nanogratings [8], or voids formation [9]. Owing strong anisotropic optical properties but also outstanding thermal properties, nanogratings have attracted significant interest in birefringent elements fabrication [10,11], microfluidic channels [12] and high temperature sensing [13,14]. Considerable research efforts have proven that, ideally (neglecting tilt effects, for example, the pulse front tilt, PFT), self-organization of the nanogratings coincides with fs laser polarization [8]. This may be attributed to a multiple scattered wave interference mechanism [15,16] and, on the other hand, illustrates that the fs laser polarization can dominantly structure the nanogratings orientation in a desirable way. Hence, polarization is a crucial parameter in determining the imprinted nanogratings distribution, which has been adopted as one dimension of the 5D data storage [17]. Within this field, Kazansky et al. recently reported a new type of fs laser induced modification which consists of random nanopores elongated perpendicularly to the laser polarization namely "Type X" modification [18] corresponding to the early birth of nanogratings. From application point of view, these nanostructures own an ultralow loss with controllable form birefringence enabled by laser polarization.

Polarization is not only important in laser materials processing, but also essential for probing and imprinting chiral structures [19,20]. Basically, an object that exhibits chirality possesses the geometric property of being incapable of coinciding with its mirror image by translations and rotations [21]. This feature predominantly results in different chiroptical responses to left- and right-handed circularly polarized light, which lays the foundations of diverse chiroptic technologies ranging from analytical chemistry to chiral switches in polarization optics [22–24]. Two decades ago, chiral gratings with double helix symmetry have been proved to own polarization-selective properties in the ground-breaking work reported by Kopp et al. [25]. Since then, chiral (or helical) long period fiber gratings (LPFGs) and fiber Bragg gratings (FBGs), fabricated on a twisted optical fiber passed through a miniature heat zone or by a helical refractive index modulation in the laser writing process, have gained more attentions in torsion and torque sensing, orbital angular momentum mode converters and circular polarizers [26–31]. The pioneered advance that a fs laser can directly induce chiral optical properties inside achiral transparent materials enriches the picture of light-matter interaction [19]. However, for now, the tailoring of fs laser generated chirality stays on 2D plane (single layer with small thickness), which restricts the level of the chiroptic effects that may be produced. 3D symmetry breaking has been proven to prominently raise the chiral optical properties compared to 2D asymmetry [32]. However, most of the existing polarization manipulations consist in 2D transformations through conventional birefringent optics (like waveplates) [33], digital holography (like Spatial Light Modulators, SLMs) [34] or geometric phase optics [35]. These strategies share a common limitation: the polarization behavior and its spatial structuration are considered only in the transverse plane, which is difficult to imprint a 3D symmetry breaking and the related chiral optical properties.

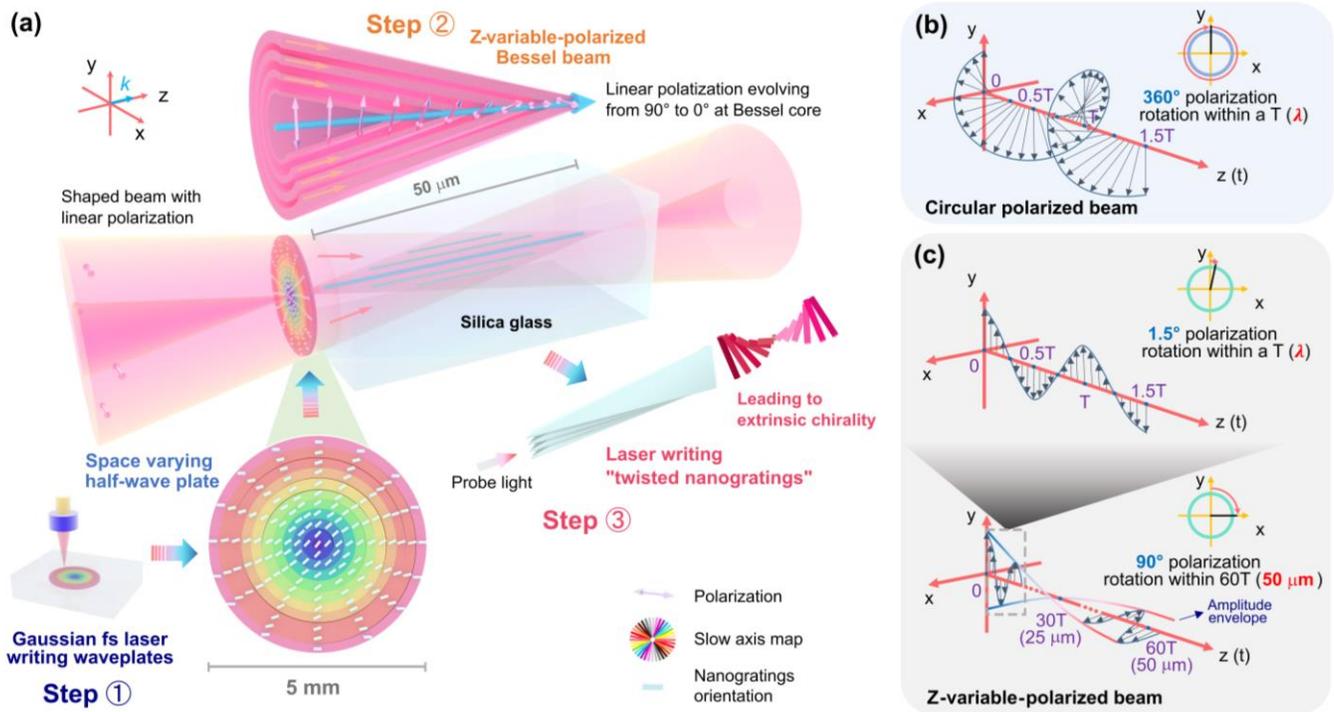

**Fig. 1.** Schematic of the concept of 3D polarization structuring of a Bessel beam via a space varying birefringent element and its application to imprint "twisted nanogratings" for extrinsic chirality. (a) Schematic of generating a z-variable-polarized fs Bessel beam by employing a space varying half-wave plate. The space varying half-wave plate is written by a Gaussian fs laser and then is inserted into the Bessel focusing path to convert a linearly polarized Bessel beam into a "z-variable-polarized" Bessel beam. Namely, it is helically polarized in the central core: the polarization state of the central lobe of the Bessel beam is linearly polarized with an orientation which evolves from 90° up to 0° within the 50 μm focus (along the optical path). Space varying half-wave plate is written at 1030 nm, 800 fs, 100 kHz, 0.16 NA, 1 mm/s, 2.5 μJ. The birefringence slow axis azimuth was varied from 0° at the outside "ring" region to 45° at the central "ring" region. Eventually, such a z-variable-polarized Bessel fs beam is used to imprint "twisted nanogratings" leading to extrinsic chirality. "$k$" is the direction of the light wavevector. (b) Schematic of the polarization distribution along the propagation of a circular polarized beam. "T" refers to the period of the light wave; "$\lambda$" refers to the wavelength of the light i.e., 800 nm for the Bessel fs laser. Here, considering light propagating in free space (speed approximates to velocity of light, c), the propagation distance of a single period T is equal to $\lambda$. (c) Schematic of the polarization distribution along the propagation of the z-variable-polarized beam within a single light wave period (above) and along the 50 μm long focus (below).

In this paper we present a unified strategy that exploits homemade space varying birefringent optical elements to structure a fs Bessel beam with a polarization function that evolves along its focus. This is a passive method for 3D polarization structuration resulting in a *z*-variable linearly polarized fs Bessel beam. Specifically, we experimentally fabricated a space varying half-wave plate by Gaussian fs laser direct writing and thus created a linearly polarized Bessel beam whose orientation is evolving along the optical propagating direction. Such a versatile longitudinal polarization shaping exhibits a strong capability of symmetry breaking along the optical path and thus, can create a strong extrinsic chirality inside transparent materials (here in silica) upon single scanning, as depicted in **Figure 1(a)**. This is just an easy-obtain but instructive example of such advanced direct writing strategy. Indeed, one can even shape the polarization in 3D for any polarization state (linear, circular or elliptical) in the different positions by adjusting the design of the space varying birefringent element. To our knowledge, this is the first report using a *z*-variable-polarized laser beam for laser-materials processing. One can envision this work not only to fabricate chiral optical elements with high efficiency, but also to inspire new ground-breaking perspectives in 3D light-matter interaction, polarization switching and light control.

## II. RESULTS

### A. Mathematic description of the z-variable-polarized Bessel beam and definitions of anisotropic optical properties

Bessel beams that possess non-diffracting nature over long distance [36] can be seen as a coherent superposition of a set of infinite plane waves with their wavevector along the generatrix of the cone half angle θ. The infinite Bessel beam is a propagation-invariant solution of the Helmholtz equation calculated by Durnin et al. [37]. Because of the conical flow of light in a Bessel beam, the polarization state of the central lobe on the optical axis at a distance z can be controlled from the plane z=0 by the polarization state of the geometrical rays emerging from a circle of radius $r = z \cdot tan(\theta)$, that will focus in (r=0, z) as shown in **Figure 1(a)**.

In our case, we choose to produce a linear polarization on the axis with a rotating orientation with propagation distance. In our concept, it can be created with an input polarization distribution where the orientation of the polarization varies with the radius at z=0 plane. (We note that the polarization state in the other lobes of the Bessel beam, outside the central core is more complex, because it is created by the interference of waves with different linear polarization orientations.)

The *x* and *y* components of the electric field vector on the optical axis can be expressed as:

$$\boldsymbol{E}(r=0,z) = E_0 \cdot \begin{bmatrix} \cos(\alpha(z)) \\ \sin(\alpha(z)) \end{bmatrix} \cdot e^{i(k_z \cdot z - \omega t)} \quad (1)$$

where $E_0$ and $k_z$ represent the on-axis Bessel beam amplitude and the longitudinal component of the wavevector, respectively. The linear polarization orientation is $\alpha(z) = \frac{\pi}{2} - \rho \cdot z$, where $\rho$ is the rotation speed parameter in rad/m. Therefore, the polarization direction evolves very slowly from 90° at the beginning of the focus (z=0 μm) to 0° at the end of the focus (z=50 μm), so $\rho = \frac{\pi/2 \, (rad)}{50 \, (\mu m)} = \pi \times 10^4 \, rad/m$ in our experiment. We have tentatively chosen a linear variation of the angle $\alpha(z)$ in our work however, we also remark that the variation can be arbitrary chosen.

The polarization state we create remains linearly polarized i.e., the field oscillates along a controlled direction, and can manipulate the orientation of laser-written nanogratings as we will see (a "twisted nanogratings structure") in "section D". This is in contrast with a circular polarization for which the polarization vector rotates in time and cannot induce such a "twisted nanogratings". It is expected because in every point, the circular polarization makes a full circle: there is zero preferential direction, as shown in **Figure 1(b)**. However, an elliptical polarized beam can imprint a strong linear birefringence due to anisotropic nanopores [38].

Whereas as depicted in **Figure 1(c)**, if we consider a same period T (corresponding to a propagation distance of $\lambda$, considering light propagating in free space namely, the propagation speed approximates to velocity of light, c), our z-variable-polarized Bessel fs beam remains linearly polarized and its orientation is evolving slowly by only 1.5°. This kind of "helical polarization distribution" can be utilized to write some chiral nanostructures (such as "twisted nanogratings" structures) that results in a significant optical chirality, as illustrated in **Figure 2(a)**.

In addition, for better describing the subsequent results section that contains many polarimetric properties, the definitions of both linear and circular optical properties are listed in **Table. 1**. Besides, it is more convenient in experiments to consider the real polarization effects caused by these anisotropic optical properties to a light beam, which are also included in the table together with their units.

Table 1. Linear and circular optical properties

| Property | Definition | Commonly described in polarimetry |
|---|---|---|
| LB (linear birefringence) | $\Delta n_L = (n_x - n_y)$ | $LB = \frac{2\pi}{\lambda}(n_x - n_y) \cdot d$ |
| LB' (45°-linear birefringence) | $\Delta n_{L'} = (n_{45°} - n_{-45°})$ | $LB' = \frac{2\pi}{\lambda}(n_{45°} - n_{-45°}) \cdot d$ |
| TLB (total linear birefringence) | $\Delta n_{TLB} = (n_e - n_o)$ <for a uniaxial material> | $TLB = \sqrt{LB^2 + LB'^2}$ |
| LD (linear dichroism) | $\Delta\kappa_L = (\kappa_x - \kappa_y)$ | $LD = \frac{2\pi}{\lambda}(\kappa_x - \kappa_y) \cdot d$ |
| LD' (45°-linear dichroism) | $\Delta\kappa_{L'} = (\kappa_{45°} - \kappa_{-45°})$ | $LD' = \frac{2\pi}{\lambda}(\kappa_{45°} - \kappa_{-45°}) \cdot d$ |
| TLD (total linear dichroism) | $\Delta\kappa_{TLD} = (\kappa_{high} - \kappa_{low})$ | $TLD = \sqrt{LD^2 + LD'^2}$ |
| CB (circular birefringence) | $\Delta n_C = (n_- - n_+)$ | $CB = \frac{2\pi}{\lambda}(n_- - n_+) \cdot d$ |
| Optical rotation | / | $\theta_r = \frac{\pi}{\lambda}(n_- - n_+) \cdot d \cdot \frac{180°}{\pi}$ |
| CD (circular dichroism) | $\Delta\kappa_C = (\kappa_- - \kappa_+)$ | $CD = \frac{2\pi}{\lambda}(\kappa_- - \kappa_+) \cdot d$ |

$n_{[\,]}$: refractive index; $\kappa_{[\,]}$: absorption index; $d$: thickness of anisotropic layer; $\lambda$: probe light wavelength.
"x": x axis; "y": y axis; "e": extraordinary light; "o": ordinary light; "-": left-handed; "+": right-handed
The table is reproduced with the permission from ref. [20]

## B. Gaussian fs laser writing space varying birefringent elements

In order to impart a 3D polarization manipulation, the key point is using a space varying birefringent element to shape, point-by-point, the state of polarization of the incident light. This 2D structured light beam with a transverse polarization shaping is then converted into a polarization variation along the optical path by using the Bessel focusing as depicted in **Figure 2(b)** resulting in an on-axis shaping i.e., "along a line".

Experimentally, this space varying birefringent element is fabricated by a Gaussian fs laser direct writing, as shown in **Figure 1(a)**. The Gaussian fs laser beam is delivered from a system (Amplitude Systèmes, Pessac, France) operating at $\lambda = 1030\ nm$, 800 fs pulse duration and 100 kHz repetition rate and then is focused 1.5 mm below the surface of a 3-mm thick silica glass wafer (Suprasil CG, Heraeus, Hanau, Germany) by an aspheric lens with a 0.16 numerical aperture (NA). Specifically, since retardance of such birefringent element is wavelength dependent [10,39], we designed and fabricated a space varying half-wave plate obtaining a $\lambda/2$ retardance for the targeted wavelength i.e., a fs Bessel beam at 800 nm. Therefore, the laser parameters were chosen to fall within nanogratings regime resulting in a controllable form birefringence. That is, the homogenous space varying half-wave plate with a diameter of 5 mm is written at 1 mm/s, 2.5 µJ/pulse with a spiral trajectory. Here, we divided the space varying half-wave plate into 10 concentric "ring" regions, which leads to a simple fabrication process. This is enough to approximate continuous shaping for the further 3D polarization manipulation. The linear polarization orientation of the Gaussian fs laser is set different in each "ring" to imprint a form birefringence with a slow axis azimuth that evolves from 0° at the outside "ring" to 45° at the central "ring". This writing process is depicted by both slow axis color map and nanogratings orientation illustration in **Figure 1(a)**.

Then, when using this homemade half-wave plate with a slow axis azimuth "rings" distribution from 0° to 45°, we obtain a linear polarization distribution of the output beam ranging from 0° outside to 90° inside, as shown in **Figure 2(c)**. This polarization distribution of the beam cross-section is measured (using a Stokes imaging polarimeter) before the geometric building of the final Bessel beam. Clearly, one can observe a well-structured radial evolution of the linear polarization orientation across the laser beam profile, except for the center. In the current design it was indeed not possible to imprint birefringence at the chosen speed at 1 mm/s within a central part of about 100 µm in diameter because this leads to a too high acceleration for our 2D xy-scanning platform. However, this technical problem can be easily overcome either by using a higher translation speed 2D scanning platform, by a scanning mirror technique or by filling the central part simply using a line-by-line scanning geometry instead of a spiral trajectory.

## C. Polarization manipulation along the optical path of a fs Bessel beam

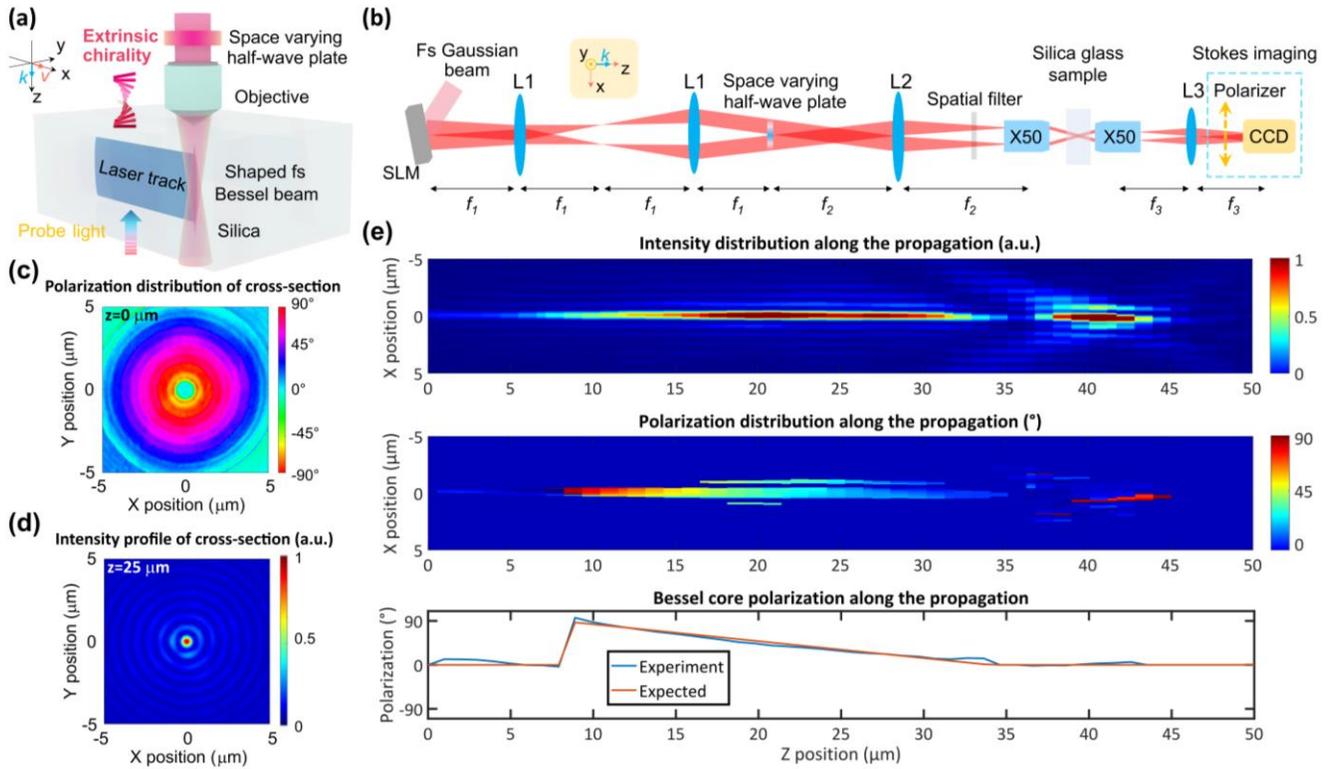

**Fig. 2.** Generation of the *z*-variable-polarized fs Bessel beam using a space varying half-wave plate. (a) Schematics of *z*-variable-polarized fs Bessel beam writing nanogratings in silica glass. *k*: the direction of the light wavevector; *v*: fs laser scanning direction. (b) Schematics of light propagation with *z*-variable polarization control (rotation). SLM: spatial light modulator; L1,2,3: focusing lenses; $f_{1,2,3}$: focal distance of the corresponding lens; ×50: objective of 50 times; CCD: charge coupled device. The dashed blue rectangle region refers to the Stokes imaging polarimeter. *k*: the direction of the light wavevector. (c) Polarization distribution (color map) of cross-section of the generated *z*-variable-polarized fs Bessel beam (recorded at z=0 μm). (d) Intensity profile of cross-section of the generated *z*-variable-polarized fs Bessel beam (recorded at z=25 μm). (e) Characterization of the generated *z*-variable-polarized fs Bessel beam along the propagation: intensity distribution (top panel), polarization distribution color map (middle panel) and Bessel core polarization distribution quantitative curve (bottom panel).

Our objective is to realize a *z*-variable-polarized fs Bessel beam that exhibits an evolving but still linear polarization state at different longitudinal *z*-axis positions and then to apply it, for the first time, to structure a chosen optical material. To validate such a "helical polarization distribution" along the focus, we firstly convert a horizontally linearly polarized (0°) Gaussian fs laser beam (from an amplified Ti: Sa laser operating at $\lambda = 800\ nm$, 120 fs pulse duration and 1 kHz repetition rate) to be a Bessel beam. This was performed by applying, using a spatial light modulator, a conical phase onto the input Gaussian beam: $\Phi(r) = -2\pi r \sin(\theta)/\lambda$, which was imaged 1:1 by a first telescope consisting of two identical lenses (L1) positioned as a classical 4*f*-configuration, as shown in **Figure 2(b)**.

Then our space varying half-wave plate was placed precisely at the relay image of the spatial light modulator to shape the 2D polarization (see **Figure 2(c)**). The polarization shaped beam was demagnified by a combination of a 750 mm lens (L2) and a ×50 microscope objective, forming a telescope with lateral magnification ×1/278. Here, a spatial filter ensures that only the zero-order Bessel beam is transmitted to the following focusing stage. Owing to the long focal region (~50 μm) of the Bessel beam, the polarization distribution in the transverse plane can be transformed so as to vary along the *z* axis, as shown conceptually in the polarization evolution schematics reported in **Figure 1(a)**. The produced *z*-variable-polarized Bessel beam has a cone half angle of $\theta = 12.5°$ (central core of 0.7 μm FWHM). In these conditions due to the use of Bessel beam, we can shape the polarization state on the axis itself i.e., along a line. Eventually, a second ×50 objective and a lens L3 are used to enable the imaging in a charge coupled device (CCD) camera. The cross-section intensity profile image of the middle of the structured light beam (z=25 μm) is recorded in **Figure 2(d)**, which well confirms the spatial profile of a Bessel beam. The intensity distribution and the polarization state distribution along the propagation of the shaped Bessel beam are also characterized experimentally, as shown in **Figure 2(e)**. Since the Bessel beam has a focal region about 50 μm, we move the CCD camera to obtain the intensity profile of the z-variable-polarized Bessel beam at each z axis position. Besides, the polarization distribution along the propagation is measured similarly moving the camera but combined with a rotating polarizer to build a Stokes imaging polarimeter [40].

It is to be mentioned that one can adjust the focusing conditions, for example, the diameter of the incident fs Gaussian beam, the apex angle of the axicon through the spatial light modulators and the telescope demagnification, so as to obtain either longer or shorter focal region of the z-variable-polarized Bessel beam. Then both the polarization spatial distribution (middle panel) and the quantitative curve (polarization of the Bessel central core, bottom panel) show a good "helical polarization distribution" after $z = 9\ \mu m$ i.e., the polarization evolves from 90° at the beginning to 0° at the end of the fs Bessel beam.

### D. Longitudinal rotation of nanogratings fabricated by the z-variable-polarized fs Bessel beam

Since the nanogratings orientation is well controlled by the fs laser polarization (ideally perpendicular, considering no PFT and normal incidence), to gain insight into their manipulation during a single laser scan, we used the z-variable-polarized fs Bessel beam to tentatively imprint "twisted nanogratings".

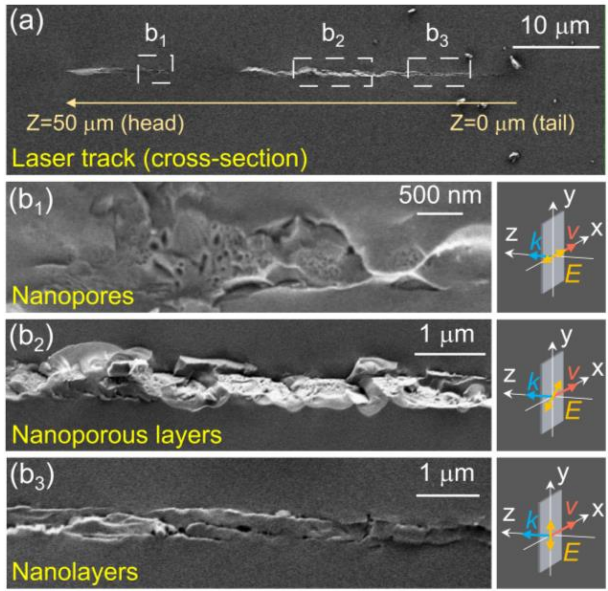

**Fig. 3**. SEM cross-section images of z-variable-polarized fs Bessel beam written nanogratings in $SiO_2$ with the parameters set at 800 nm, 120 fs, 1 kHz, 0.8 NA, 10 μm/s, 0.6 μJ. (a) An overall view of the cleaved laser track. Detailed pictures of the head ($b_1$), the middle ($b_2$) and the tail ($b_3$) of the laser track. The axial labels are presented at the right part correspond to each detailed SEM images, respectively. $k$: the direction of the light wavevector; $v$: fs laser scanning direction; $E$: the electric field direction of the fs Bessel beam (polarization direction).

Experimentally, the laser track is written by scanning the z-variable-polarized fs Bessel beam along x axis inside the silica glass, as depicted in **Figure 2(a)**. Then the written line is cut perpendicular to the x axis to get the cross-section of the laser track which is in y-z plane. Thus, we are able to see the nanogratings orientation rotation along z axis in single laser track, as shown in "Step ③" of **Figure 1(a)**. This schematic image shows the cross-section view of the fs laser written track where the nanogratings orientation is rotating. Note that this kind of "twisted nanogratings" is obtained in one single structure written by z-variable-polarized fs Bessel beam single scanning and is then captured by SEM (Field-Emission Gun Scanning Electron Microscope, ZEISS SUPRA 55 VP, 1 kV accelerating voltage) using secondary electrons imaging as seen in **Figure 3**. Obviously, the SEM images shows a rotation morphology from head to tail of the laser-written track due to the linear polarization evolving along the fs Bessel laser beam. **Figure 3(a)** depicts the global view of the z-variable-polarized fs Bessel beam induced track. The head, middle and tail regions of the laser track are zoomed in **Figures 3($b_1$), 3($b_2$) and 3($b_3$)**, respectively. Note that because the polarization rotation plane is perpendicular to the SEM observation cross-section (making it difficult to describe the polarization rotation in a 2D drawing), the axial labels on the right of each detailed SEM images are slightly tilted as to depict the different polarization directions at different positions from a 3D viewpoint. Here, in order to describe the SEM images related to the nanogratings distribution, we define the polarization configuration as "scanning direction + polarization direction" since the laser polarization is different at different z positions of the z-variable-polarized fs Bessel beam.

One can observe clear nanopores at the head of the laser track due to an "Xx" polarization configuration and well oriented nanolayers at the tail of the laser track due to an "Xy" polarization configuration. In the middle of the laser track, one can see the nanoporous layers due to a "X+45°" configuration since the layers are tilted out of the plane. The depicted nanopores located in the nanolayer are mainly attributed to the silica decomposition from $SiO_2$ to $SiO_{2(1-x)} + x \cdot O_2$. This subsequently results in an expulsion of ionized oxygen atoms to the surrounding lattice assisted by a tensile stress, which has been recently revisited as a tensile stress assisted nano-cavitation process leading to the formation of an "self-assembly" of anisotropic nanopores [41,42].

The experimental results show a kind of "twisted nanogratings" distribution. It is thus a fact that the z-variable-polarized fs Bessel beam possesses different linearly polarization orientations at different z positions leading to different arrangements of the nanogratings (from head to tail). Hence, it can be expected that an arbitrary longitudinal orientation distribution of nanogratings can be enabled by designed polarization conversions of the laser beam through space varying birefringent elements, combined with Bessel focusing.

### E. Z-variable-polarized fs Bessel beam written chirality

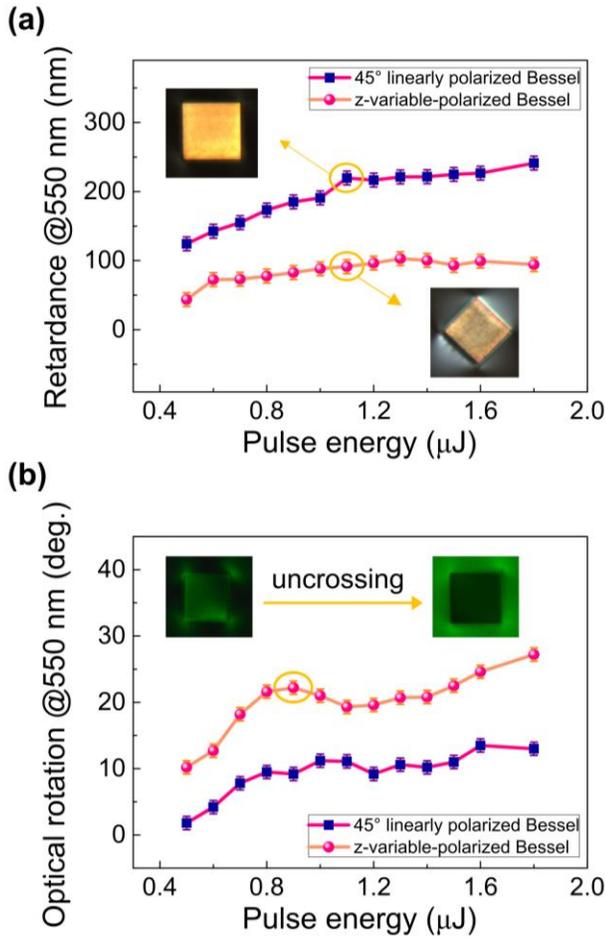

**Fig. 4**. Optical characterizations enabled by crossed-polarizer microscopy. SiO$_2$ samples are written at 800 nm, 120 fs, 1 kHz, 0.8 NA, 10 μm/s. (a) Retardance measurements at different laser pulse energies written by both the 45° linearly polarized fs Bessel beam and the *z*-variable-polarized fs Bessel beam. (b) Optical rotation measurements with different laser pulse energies written by both the 45° linearly polarized fs Bessel beam and the *z*-variable-polarized fs Bessel beam. Measurements are done at 550 nm.

As a result of the varying orientation (here, a 3D rotation) of the nanogratings generated by the *z*-variable-polarized fs Bessel beam, strong symmetry breaking is obtained, leading to chiral optical properties. To highlight the achieved optical performance, a 45° linearly polarized fs Bessel beam was used for comparison. The latter choice is justified because 45° linearly polarized fs laser beams have been proven to imprint highest chiral optical properties among all linearly polarized fs laser beams [19,20].

First, the retardance induced by both the 45° linearly polarized fs Bessel beam and the *z*-variable-polarized fs Bessel beam was characterized by a polarizing microscope (Olympus BX51, Tokyo, Japan) equipped with a "de Sénarmont" compensator. The "de Sénarmont" arrangement uses a highly precise quarter-wave plate and a 180-degree-rotation analyzing polarizer to perform retardance measurements with a typical error range within ±10 nm at the wavelength of 550 nm. Here, the retardance is the optical phase difference related to TLB; the relationship between these two quantities is given by $TLB/\pi = retardance/\lambda$.

**Figure 4(a)** reports the retardance results of samples irradiated by both 45° linearly polarized fs Bessel beam and *z*-variable-polarized fs Bessel beam at different pulse energies. Clearly, the retardance induced by the *z*-variable-polarized fs Bessel beam is less (by one half roughly) than that induced by the 45° linearly polarized fs Bessel beam. It is mainly attributed to the (partial) suppression of linear birefringence due to the different slow axes at different *z* positions, as investigated in our previous work [33]. In addition, the insets show the corresponding crossed-polarizer microscopic images of the irradiated samples at the pulse energy of 1.1 μJ as an illustration. The colors of different samples are due to different retardance amplitudes in agreement with Michel-Levy color chart which also confirms a lower retardance when using the *z*-variable-polarized fs Bessel beam instead of the 45° linearly polarized fs Bessel beam. Further, since an object that possesses chiral optical properties can rotate the polarization plane of a linearly polarized probe light (a so-called optical rotation), one can measure this optical rotation (with an error of ±1°) to unveil the chiral optical properties in a simple way as shown in **Figure 4(b)**. The results clearly indicate higher (twice as high, roughly) optical rotation values for the *z*-variable-polarized fs Bessel beam written samples compared to the 45° linearly polarized fs Bessel beam one.

As confirmed experimentally, the *z*-variable-polarized fs Bessel beam can modify the nanogratings orientation along the optical path, thus forming a chiral structure on the micrometric scale. This appealing feature allows the fabrication of chiral waveplates (waveplates that exploit chiral optical properties to control the light propagation, especially for polarization manipulation) with high circular birefringence and a quite small (ideally none) linear birefringence. To quantitatively characterize the performances of imprinting extrinsic chirality in SiO$_2$ using this *z*-variable-polarized fs Bessel beam, a spectroscopic Mueller polarimeter (based on a modified Smart SE ellipsometer, JY HORIBA) was used to obtain the Mueller matrix of each sample, which was subsequently decomposed by the differential decomposition, thus providing the differential matrix of the irradiated sample [43,44], which allows to easily determine all the anisotropic optical properties of the sample as follows:

$$M_{diff} = \begin{bmatrix} 0 & LD & LD' & CD \\ LD & 0 & CB & -LB' \\ LD' & -CB & 0 & LB \\ CD & LB' & -LB & 0 \end{bmatrix} \quad (2)$$

As a result, the values (expressed in radians) of both the linear properties and the circular properties are obtained simultaneously. The TLB value can be calculated by $TLB = \sqrt{LB^2 + LB'^2}$, as shown in **Table 1**.

Experimentally, homogeneous square-shaped samples with a size of 0.2 mm ×0.2 mm were written using either *z*-variable-polarized fs Bessel beam, 45° linearly polarized fs Bessel beam and 0° linearly polarized fs Gaussian beam (for comparison). Each square is made of a set of lines with a pitch of 1 μm and the line scanning direction is always along "+X" direction to avoid any non-reciprocal or "quill" writing [45,46]. Then the samples were measured with the spectroscopic Mueller polarimeter in transmission mode. **Figure 5** displays the spectroscopic results for both TLB (solid lines) and CB (dashed lines) of a z-variable-polarized fs Bessel beam (red), a 45° linearly polarized fs Bessel beam (blue) and a 0° linearly polarized fs Gaussian beam (gray). A large difference in both TLB and CB values can be observed between the different fs laser beam configurations. Clearly, the z-variable-polarized fs Bessel beam simultaneously induces large CB yet small TLB over the entire measuring wavelength range from 450 nm to 1000 nm. By the way, both anisotropic optical properties exhibit large values at small wavelengths and decrease with increasing the wavelength due to their 1/λ inherent behavior according to the common descriptions depicted in **Table 1**.

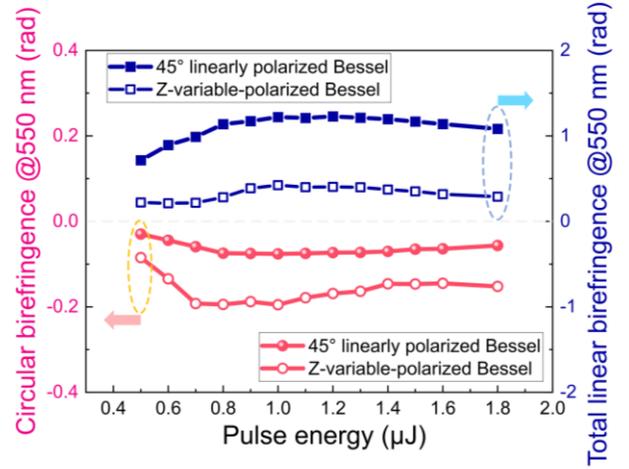

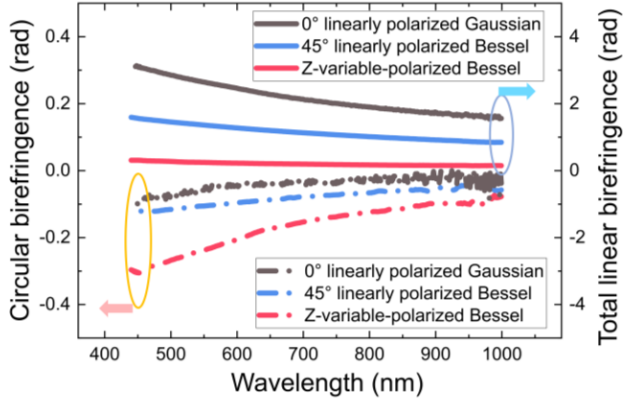

**Fig. 5**. Spectroscopic Mueller polarimetry results for circular birefringence (dashed lines) and total linear birefringence (solid lines) of SiO$_2$ samples written by a 0° linearly polarized fs Gaussian beam (gray), a 45° linearly polarized fs Bessel beam (blue) and a z-variable-polarized fs Bessel beam (red). Laser parameters are 800 nm, 120 fs, 1 kHz, 0.8 NA, 10 μm/s, 0.7 μJ.

To further explore the optical response and to tailor the optical properties of the z-variable-polarized fs Bessel beam written extrinsic chirality in silica glass, a set of pulse energies (all the other laser parameters being fixed) was adopted for investigating both TLB and CB properties, as shown in **Figure 6**. Thus, for the z-variable-polarized fs Bessel beam written samples, the CB value reaches −0.2 rad (which, in absolute value, is three times as much as the CB created by the 45° linearly polarized fs Bessel beam) over the laser pulse energy range of 0.7-1.0 μJ. Considering the simultaneously induced TLB, 0.7 μJ is definitely the optimal laser energy parameter since it exhibits a TLB value of only 0.2 rad, which is four times as less as the TLB induced by the 45° linearly polarized fs Bessel beam. In similar conditions, a 0° linearly polarized fs Bessel (or Gaussian) beam writing would generate negligible circular properties [20].

**Fig. 6**. Mueller polarimetry results of circular birefringence (red) and total linear birefringence (blue) of the SiO$_2$ samples written at different laser pulse energies by both 45° linearly polarized fs Bessel beam (full symbols) and z-variable-polarized fs Bessel beam (hollow symbols). Measurements are done at 550 nm. SiO$_2$ samples are written at 800 nm, 120 fs, 1 kHz, 0.8 NA, 10 μm/s.

These results explicitly provide a new pathway to fabricate chiral waveplates with high CB and almost suppressed TLB. This is a novel and more integrated approach, which differs from our previous strategy of utilizing multilayer configurations [33].

## III. DISCUSSION AND CONCLUSION

*Controlling the nanogratings orientation by fs laser polarization:* Since the first morphology observations performed by Shimotsuma et al. in 2003 [8], fs laser written nanogratings have attracted tremendous attention in the ultrafast light-matter interaction research field. However, until now, the mechanism of nanogratings formation still remains under discussion. Several compelling assumptions emerge in explaining the fundamentals of the nanogratings formation process. For instance, interference between the light field and a bulk plasma wave may account for the nanogratings formation, especially in explaining their periodicity, as proposed by Kazansky et al. [8] whereas other groups proposed a scattered waves interference model [15,16]. Recently, our group has revealed the occurrence of elongated nanopores thus forming the nanolayers. Ultrafast silica decomposition through nano-cavitation process was proposed as a tentative explanation [41]. Although a direct evidence for the ultrafast process is difficult to be obtained because of the limited frame rate and time resolution of camera, the imprinted orientation of nanogratings distribution has been proved to be almost perpendicular to the laser polarization [8]. In addition, it has been discussed that PFT can lead to a small tilt in the nanogratings orientation due to spatiotemporal chirp and angular dispersion. When the light propagates to the focal plane, PFT-related effects grow exponentially and can induce a few degrees of tilt at 100 fs/mm to the nanogratings [47].

Nevertheless, this tilt in question represents only a small deviation from the nanogratings orientation.

*Origin of the written chirality:* The chirality originates from molecular chiral arrangement at nanoscale, that is, intrinsic chirality contributes to optical activity, and thus, to the manifestation of chiral optical properties. However, the chiral optical properties can also be due to the so called extrinsic chirality which is related to a chiral object constructed in an experimental arrangement not coincident with its mirror image [48]. Hence, recently we have advanced a novel conceptual view according to which the fs laser written extrinsic chirality arises from two contributions: a form-related birefringence and a stress-related birefringence with non-parallel and non-perpendicular neutral axes. We therefore deal with extrinsic chirality arising from the experimental arrangement of linear optical properties [20]. Compared to intrinsic chirality, extrinsic chirality can exhibit a larger optical rotation due to the strong symmetry breaking at a larger, typically micrometric scale. In this paper, the *z*-variable-polarized fs Bessel beam driven orientation rotated nanogratings structure is a form of experimental chiral arrangement corresponding to extrinsic chirality and is potentially useful in the fabrication of chiral devices with large and controllable optical rotation.

*Z-variable-polarized fs Bessel beam writing vs multilayer strategy:* Concerning extrinsic chirality implantation, the multilayer strategy has proven to be a realizable approach in generating high CB with minimal TLB due to the (partial) suppression of linear birefringence through the alignment of fast/slow axes of different layers [20,33]. However, such a multilayer strategy features limitations such as low fabrication speed and requirement for precise alignment between layers. Geometrically, the multilayer strategy amounts to fabricating a staircase-like chiral arrangement without transition region between layers. Conversely, the *z*-variable-polarized fs Bessel beam writing approach can (1) perform the writing task in a single scan, which provides high efficiency and thus, unlocks the versatility of fabricating chiral devices in a fast way; (2) more importantly, *z*-variable-polarized fs Bessel beam irradiation enables a quasi-gradual evolution of polarization, that is, a smoother variation of the nanogratings orientation compared to the staircase-like structure written by using the multilayer strategy. Note that by further engineering the space varying birefringent elements, one can tailor a continuous chiral arrangement with choosing the pitch and the length of the "twisted silica glass". However, the absolute CB value appears not so large if compared, for example, to our best multilayer structure results (CB=-2.25 rad with a total thickness of 340 μm, measured at 550 nm) [33]. Therefore, for a multilayer strategy, the CB per unit length is calculated to be -6.62 rad/mm whereas for a *z*-variable-polarized fs Bessel beam writing it is around -5 rad/mm (i.e., -0.2 rad with a thickness of 40 μm). This preliminary result is attributed to the relatively short extent of the Bessel beam used in this study, which could be enlarged by for example, using a smaller demagnification factor of the imaging telescope. Nevertheless, if compared to 45° linearly polarized fs Bessel beam writing under identical conditions, our results clearly prove the potential of implanting high CB by using such kind of controllable "helical polarization distribution".

Towards practical applications, one needs to consider the limitations of this strategy as well: (1) To date, the fs laser written space varying birefringent elements are based on self-organized nanogratings whose nanoporous structures can lead to strong scattering losses. The spectral signature possesses a $1/\lambda^4$ dependence and corresponds to a strong anisotropic loss in ultraviolet-visible range [49], which exemplified to be a 74% transmittance measured at 550 nm. However, this restriction is expected to be partly solved by employing a "Type X" modification with its ultralow loss [18]. (2) It is difficult to use our shaping strategy in mid-Infrared range that needs further work to imprint higher birefringence such as in ZnO-Barium Gallo-Germanate glasses [50]. (3) The spatial resolution of the fs written space varying waveplate is on the order of micrometer size, which might restrict the potentials of applications in sub-micrometer and nanometer scale integrated systems.

*Conclusion and potential applications:* In summary, in the sight of the current state of light control, we proposed a novel concept to extend the polarization manipulation along the optical path, which provides new perspectives in polarization transformation and therefore, enriches the state of the art of light manipulation. A stable "helical polarization distribution" fs Bessel beam was generated for the first time by using a fs written space varying half-wave plate. Besides, for the first time at our knowledge such a *z*-variable-polarized laser beam was implemented for laser materials processing namely in silica glass. This enables the experimental achievement of twice as large circular birefringence with, simultaneously, four times smaller linear birefringence, compared to a 45° linearly polarized fs Bessel beam or Gaussian beam.

Several applications would benefit from this 3D polarization structuring strategy: (1) this technique can be directly used for chiral waveplates fabrication towards linear polarization rotators with high optical rotation and ultimately no linear birefringence. This kind of optical rotators own advantages of alignment-free compared to conventional half-wave plate and potential achromatic properties. (2) Since polarization is an essential parameter in 5D data storage [5], our strategy of 3D control of polarization may be useful for increasing the writing and reading speed. The optical chirality could even be a new dimension for increasing the capacity of the information storage. (3) This 3D polarization structuring approach will benefit the fundamental understanding of light-matter interaction and will provide ultrafast light manipulation with additional degrees of freedom. For example, this could provide a platform for laser-based production of cholesteric liquid crystal analogous optical devices using tiny lengths of inorganic glass i.e., "twisted silica glass". In general, it will not only facilitate the implementation of polarization optics through the mediation of space varying optical devices but also underpin numerous potential chiroptic applications such as passive isolators, quantum emitters, optical tweezers, achromatic devices and beyond.

## ACKNOWLEDGMENTS

This work is supported by Agence Nationale pour la Recherche, FLAG-IR project (ANR-18-CE08-0004-01) and European Research Council (ERC) 682032-PULSAR; Jiafeng Lu acknowledges the China Scholarship Council (CSC) for the funding No. 202006890077.

## AUTHOR DECLARATIONS

### Conflict of Interest
The authors have no conflicts to disclose.

### Author Contributions
M.L. supervised the project. J.L., F.C. and M.L. wrote the manuscript. J.L. carried out the design and fabrication of the space varying birefringent elements. M.H. and F.C. carried out the Bessel beam generation and laser writing. J.L. carried out the sample treatment and optical properties characterization. J.L. and E.G.-C. performed the Mueller polarimetry measurements. J.L., F.B. and M.L. performed the SEM measurements. J.L., M.H., F.C. and M.L. processed and analyzed the data. J.L., M.L., F.C., E.G.-C., R.O., B.P. and X.Z. discussed the experimental results. All authors commented and discussed this work.

## DATA AVAILABILITY

The data that support the findings of this study are available from the corresponding author upon reasonable request.

## References

1. K. Sugioka and Y. Cheng, "Ultrafast lasers-reliable tools for advanced materials processing," Light Sci. Appl. **3**(1), e149 (2014).
2. D. Wen, J. J. Cadusch, J. Meng, and K. B. Crozier, "Vectorial holograms with spatially continuous polarization distributions," Nano Lett. **21**(4), 1735–1741 (2021).
3. C. Xie, R. Meyer, L. Froehly, R. Giust, and F. Courvoisier, "In-situ diagnostic of femtosecond laser probe pulses for high resolution ultrafast imaging," Light Sci. Appl. **10**(1), 126 (2021).
4. A. Yan, S. Huang, S. Li, R. Chen, P. Ohodnicki, M. Buric, S. Lee, M. J. Li, and K. P. Chen, "Distributed Optical Fiber Sensors with Ultrafast Laser Enhanced Rayleigh Backscattering Profiles for Real-Time Monitoring of Solid Oxide Fuel Cell Operations," Sci. Rep. **7**(1), 9360 (2017).
5. H. Wang, Y. Lei, L. Wang, M. Sakakura, Y. Yu, G. Shayeganrad, and P. G. Kazansky, "100-Layer Error-Free 5D Optical Data Storage by Ultrafast Laser Nanostructuring in Glass," Laser Photonics Rev. **16**(4), 2100563 (2022).
6. M. Beresna, M. Gecevičius, and P. G. Kazansky, "Ultrafast laser direct writing and nanostructuring in transparent materials," Adv. Opt. Photonics **6**(3), 293–339 (2014).
7. K. M. Davis, K. Miura, N. Sugimoto, and K. Hirao, "Writing waveguides in glass with a femtosecond laser," Opt. Lett. **21**(21), 1729–1731 (1996).
8. Y. Shimotsuma, P. G. Kazansky, J. Qiu, and K. Hirao, "Self-organized nanogratings in glass irradiated by ultrashort light pulses," Phys. Rev. Lett. **91**(24), 247405 (2003).
9. X. Wang, F. Chen, Q. Yang, H. Liu, H. Bian, J. Si, and X. Hou, "Fabrication of quasi-periodic micro-voids in fused silica by single femtosecond laser pulse," Appl. Phys. A Mater. Sci. Process. **102**(1), 39–44 (2011).
10. M. Lancry, R. Desmarchelier, K. Cook, B. Poumellec, and J. Canning, "Compact birefringent waveplates photo-induced in silica by femtosecond laser," Micromachines **5**(4), 825–838 (2014).
11. J. Lu, Y. Dai, Q. Li, Y. Zhang, C. Wang, F. Pang, T. Wang, and X. Zeng, "Fiber nanogratings induced by femtosecond pulse laser direct writing for in-line polarizer," Nanoscale **11**(3), 908–914 (2019).
12. C. Hnatovsky, R. S. Taylor, E. Simova, P. P. Rajeev, D. M. Rayner, V. R. Bhardwaj, and P. B. Corkum, "Fabrication of microchannels in glass using focused femtosecond laser radiation and selective chemical etching," Appl. Phys. A Mater. Sci. Process. **84**(1–2), 47–61 (2006).
13. S. J. Mihailov, D. Grobnic, C. Hnatovsky, R. B. Walker, P. Lu, D. Coulas, and H. Ding, "Extreme environment sensing using femtosecond laser-inscribed fiber bragg gratings," Sensors **17**(12), 2909 (2017).
14. Y. Wang, M. Cavillon, J. Ballato, T. Hawkins, T. Elsmann, M. Rothhardt, R. Desmarchelier, G. Laffont, B. Poumellec, and M. Lancry, "3D Laser Engineering of Molten Core Optical Fibers: Toward a New Generation of Harsh Environment Sensing Devices," Adv. Opt. Mater. **10**(18), 2200379 (2022).
15. R. Buschlinger, S. Nolte, and U. Peschel, "Self-organized pattern formation in laser-induced multiphoton ionization," Phys. Rev. B **89**(18), 184306 (2014).
16. A. Rudenko, J. P. Colombier, T. E. Itina, and R. Stoian, "Genesis of Nanogratings in Silica Bulk via Multipulse Interplay of Ultrafast Photo-Excitation and Hydrodynamics," Adv. Opt. Mater. **9**(20), 2100973 (2021).
17. J. Zhang, M. Gecevičius, M. Beresna, and P. G. Kazansky, "Seemingly unlimited lifetime data storage in nanostructured glass," Phys. Rev. Lett. **112**(3), (2014).
18. M. Sakakura, Y. Lei, L. Wang, Y. H. Yu, and P. G. Kazansky, "Ultralow-loss geometric phase and polarization shaping by ultrafast laser writing in silica glass," Light Sci. Appl. **9**(1), 15 (2020).
19. B. Poumellec, M. Lancry, R. Desmarchelier, E. Hervé, and B. Bourguignon, "Parity violation in chiral structure creation under femtosecond laser irradiation in silica glass?," Light Sci. Appl. **5**(11), e16178 (2016).
20. J. Lu, J. Tian, B. Poumellec, E. Garcia-caurel, R. Ossikovski, X. Zeng, and M. Lancry, "Tailoring chiral optical properties by femtosecond laser direct writing in silica," Light Sci. Appl. **12**(1), 46 (2023).
21. Y. Kim, B. Yeom, O. Arteaga, S. J. Yoo, S. G. Lee, J. G. Kim, and N. A. Kotov, "Reconfigurable chiroptical nanocomposites with chirality transfer from the macro- to the nanoscale," Nat. Mater. **15**(4), 461–468 (2016).
22. M. B. Steffensen, D. Rotem, and H. Bayley, "Single-molecule analysis of chirality in a multicomponent reaction network," Nat. Chem. **6**(7), 603–607 (2014).

23. Y. Wang and Q. Li, "Light-driven chiral molecular switches or motors in liquid crystals," Adv. Mater. **24**(15), 1926–1945 (2012).
24. J. K. Gansel, M. Thiel, M. S. Rill, M. Decker, K. Bade, V. Saile, G. Von Freymann, S. Linden, and M. Wegener, "Gold helix photonic metamaterial as broadband circular polarizer," Science. **325**(5947), 1513–1515 (2009).
25. V. I. Kopp, V. M. Churikov, J. Singer, N. Chao, D. Neugroschl, and A. Z. Genack, "Chiral Fiber Gratings," Science (80-. ). **305**(5680), 74–75 (2004).
26. V. I. Kopp and A. Z. Genack, "Chiral fibres: Adding twist," Nat. Photonics **5**(8), 470–472 (2011).
27. C. Liao, K. Yang, J. Wang, Z. Bai, Z. Gan, and Y. Wang, "Helical Microfiber Bragg Grating Printed by Femtosecond Laser for Refractive Index Sensing," IEEE Photonics Technol. Lett. **31**(12), 971–974 (2019).
28. S. Oh, K. R. Lee, U.-C. Paek, and Y. Chung, "Fabrication of helical long-period fiber gratings by use of a $CO_2$ laser," Opt. Lett. **29**(13), 1464–1466 (2004).
29. C. Ma, J. Wang, and L. Yuan, "Review of helical long-period fiber gratings," Photonics **8**(6), 1–22 (2021).
30. X. Zhang, A. Wang, R. Chen, Y. Zhou, H. Ming, and Q. Zhan, "Generation and Conversion of Higher Order Optical Vortices in Optical Fiber with Helical Fiber Bragg Gratings," J. Light. Technol. **34**(10), 2413–2418 (2016).
31. J. Canning, Y. Wang, M. Lancry, Y. Luo, and G.-D. Peng, "Helical distributed feedback fiber Bragg gratings and rocking filters in a 3D printed preform-drawn fiber," Opt. Lett. **45**(19), 5444 (2020).
32. O. Arteaga, J. Sancho-Parramon, S. Nichols, B. M. Maoz, A. Canillas, S. Bosch, G. Markovich, and B. Kahr, "Relation between 2D/3D chirality and the appearance of chiroptical effects in real nanostructures," Opt. Express **24**(3), 2242–2252 (2016).
33. J. Lu, E. Garcia-Caurel, R. Ossikovski, F. Courvoisier, X. Zeng, B. Poumellec, and M. Lancry, "Femtosecond laser direct writing multilayer chiral waveplates with minimal linear birefringence," Opt. Lett. **48**(2), 271–274 (2023).
34. S. Hasegawa, K. Shiono, and Y. Hayasaki, "Femtosecond laser processing with a holographic line-shaped beam," Opt. Express **23**(18), 23185–23194 (2015).
35. C. P. Jisha, S. Nolte, and A. Alberucci, "Geometric Phase in Optics: From Wavefront Manipulation to Waveguiding," Laser Photonics Rev. **15**(10), 2100003 (2021).
36. M. K. Bhuyan, F. Courvoisier, P. A. Lacourt, M. Jacquot, R. Salut, L. Furfaro, and J. M. Dudley, "High aspect ratio nanochannel machining using single shot femtosecond Bessel beams," Appl. Phys. Lett. **97**(8), 081102 (2010).
37. J. Durnin, J. Miceli, and J. H. Eberly, "Diffraction-free beams," Phys. Rev. Lett. **58**(15), 1499–1501 (1987).
38. Y. Lei, G. Shayeganrad, H. Wang, C. Deng, and P. G. Kazansky, "Towards efficient nanostructuring of silica glass by elliptically polarized ultrafast laser pulses," in *Conference on Lasers and Electro-Optics, Technical Digest Series (Optica Publishing Group)* (2022), p. SF3L.4.
39. M. Beresna, M. Gecevičius, P. G. Kazansky, and T. Gertus, "Radially polarized optical vortex converter created by femtosecond laser nanostructuring of glass," Appl. Phys. Lett. **98**(20), (2011).
40. G. Milione, A. Dudley, T. A. Nguyen, O. Chakraborty, E. Karimi, A. Forbes, and R. R. Alfano, "Experimental measurement of the self-healing of the spatially inhomogeneous states of polarization of radially and azimuthally polarized vector Bessel beams," Print at https://doi.org/10.48550/arXiv.1412.2722 (2014).
41. M. Lancry, B. Poumellec, J. Canning, K. Cook, J. C. Poulin, and F. Brisset, "Ultrafast nanoporous silica formation driven by femtosecond laser irradiation," Laser Photonics Rev. **7**(6), 953–962 (2013).
42. M. Lancry, J. Canning, K. Cook, M. Heili, D. R. Neuville, and B. Poumellec, "Nanoscale femtosecond laser milling and control of nanoporosity in the normal and anomalous regimes of GeO2-SiO2 glasses," Opt. Mater. Express **6**(2), 321–330 (2016).
43. R. Ossikovski, "Differential matrix formalism for depolarizing anisotropic media," Opt. Lett. **36**(12), 2330–2332 (2011).
44. R. Ossikovski and O. Arteaga, "Statistical meaning of the differential Mueller matrix of depolarizing homogeneous media," Opt. Lett. **39**(15), 4470–4473 (2014).
45. B. Poumellec, M. Lancry, R. Desmarchelier, E. Hervé, F. Brisset, and J. C. Poulin, "Asymmetric Orientational Writing in glass with femtosecond laser irradiation," Opt. Mater. Express **3**(10), 1586–1599 (2013).
46. P. G. Kazansky, W. Yang, E. Bricchi, J. Bovatsek, A. Arai, Y. Shimotsuma, K. Miura, and K. Hirao, ""Quill" writing with ultrashort light pulses in transparent materials," Appl. Phys. Lett. **90**(15), 12–15 (2007).
47. Y. Dai, J. Ye, M. Gong, X. Ye, X. Yan, G. Ma, and J. Qiu, "Forced rotation of nanograting in glass by pulse-front tilted femtosecond laser direct writing," Opt. Express **22**(23), 28500–28505 (2014).
48. E. Plum, "Extrinsic chirality: Tunable optically active reflectors and perfect absorbers," Appl. Phys. Lett. **108**(24), 241905 (2016).
49. M. Beresna, M. Gecevičius, M. Lancry, B. Poumellec, and P. G. Kazansky, "Broadband anisotropy of femtosecond laser induced nanogratings in fused silica," Appl. Phys. Lett. **103**(13), 131903 (2013).
50. H. Yao, R. Zaiter, M. Cavillon, P. Delullier, B. Lu, T. Cardinal, Y. Dai, B. Poumellec, and M. Lancry, "Formation of nanogratings driven by ultrafast laser irradiation in mid-IR heavy oxide glasses," Ceram. Int. **48**(21), 31363–31369 (2022).